\documentclass[prd,showkeys,floatfix,twocolumn,amsmath,amssymb,floatfix]{revtex4}
\usepackage{graphicx}
\usepackage{epstopdf}
\usepackage{multirow}
\usepackage{subfigure}
\usepackage[colorlinks,citecolor=blue,anchorcolor=red,menucolor=red,linkcolor=red,filecolor=red,runcolor=red,urlcolor=blue,frenchlinks=red]{hyperref}
\usepackage{enumitem}

\newcommand{\feynp}[1]{#1\kern-0.45em/}

\def\FF(s){\left[(\alpha+\beta)m_c^2-\alpha\beta s\right]}
\def\HH(s){\left[m_c^2-\alpha(1-\alpha) s\right]}
\def\KK(s){\left[\gamma m_c^2-\gamma(1-\gamma) s\right]}
\def\non{\\ \nonumber}

\makeatletter
\renewcommand{\@thesubfigure}{\hskip\subfiglabelskip}
\makeatother
\allowdisplaybreaks[3]

\begin{document}
\title{Double-gluon charmonium hybrid states with various (exotic) quantum numbers}
%

\author{Niu Su$^{1,2}$}
\author{Hua-Xing Chen$^1$}
\email{hxchen@seu.edu.cn}
\author{Wei Chen$^3$}
\email{chenwei29@mail.sysu.edu.cn}
\author{Shi-Lin Zhu$^4$}
\email{zhusl@pku.edu.cn}

\affiliation{
$^1$School of Physics, Southeast University, Nanjing 210094, China\\
$^2$Research Center for Nuclear Physics (RCNP), Osaka University, Ibaraki 567-0047, Japan\\
$^3$School of Physics, Sun Yat-Sen University, Guangzhou 510275, China\\
$^4$School of Physics and Center of High Energy Physics, Peking University, Beijing 100871, China}
\begin{abstract}
We study the double-gluon charmonium hybrid states with various quantum numbers, each of which is composed of one valence charm quark and one valence charm antiquark as well as two valence gluons. We concentrate on the exotic quantum numbers $J^{PC} =0^{--}/0^{+-}/1^{-+}/2^{+-}/3^{-+}$ that the conventional $\bar q q$ mesons can not reach. We apply the QCD sum rule method to calculate their masses to be $7.28^{+0.38}_{-0.43}$~GeV, $5.19^{+0.36}_{-0.46}$~GeV, $5.46^{+0.41}_{-0.62}$~GeV, $4.48^{+0.25}_{-0.31}$~GeV, and $5.54^{+0.35}_{-0.43}$~GeV, respectively. We study their possible decay patterns and propose to search for the $J^{PC}=2^{+-}/3^{-+}$ states in the $D^*\bar D^{(*)}/D^{*}_s \bar D^{(*)}_s/\Sigma_c^* \bar \Sigma_c^{(*)}/\Xi_c^* \bar \Xi_c^{(\prime,*)}$ channels. Experimental investigations on these states and decay channels can be useful in classifying the nature of the hybrid state, thus serving as a direct test of QCD in the low energy sector.
\end{abstract}
\keywords{hybrid state, exotic hadron, exotic quantum number, QCD sum rules}
\date{\today}

\maketitle

\vspace{2cm}

{\it Introduction} ---
A hybrid state is composed of one valence quark and one valence antiquark as well as one or more valence gluons. Especially, the hybrid states with $J^{PC} =0^{--}/0^{+-}/1^{-+}/2^{+-}/3^{-+}/\cdots$ are of particular interests, since these exotic quantum numbers can not be reached by the conventional $\bar q q$ mesons~\cite{pdg}. Up to now there are four structures observed in experiments with the exotic quantum number $J^{PC} = 1^{-+}$, {\it i.e.}, the $\pi_1(1400)$~\cite{IHEP-Brussels-LosAlamos-AnnecyLAPP:1988iqi}, $\pi_1(1600)$~\cite{E852:1998mbq}, $\pi_1(2015)$~\cite{E852:2004gpn}, and $\eta_1(1855)$~\cite{BESIII:2022riz}. They are good candidates for the single-gluon hybrid states that contain only one valence gluon, while they may also be explained as the compact tetraquark states or hadronic molecular states~\cite{Chen:2008ne,Zhang:2019ykd,Dong:2022cuw,Wan:2022xkx,Wang:2022sib}. In the past half century there have been a lot of experimental and theoretical investigations on these hybrid states~\cite{Barnes:1977hg,Balitsky:1982ps,Frere:1988ac,Close:1994hc,Lacock:1996ny,MILC:1997usn,Chetyrkin:2000tj,Jin:2002rw,Dudek:2009qf,Chen:2010ic,Chen:2022isv,Qiu:2022ktc}. However, their nature still remains elusive, partly due to the difficulty in differentiating the hybrid and multiquark pictures~\cite{Chen:2022asf,Klempt:2007cp,Meyer:2015eta,Ketzer:2019wmd}. This tough problem needs to be solved in future by experimentalists and theorists together.

In this letter we investigate the double-gluon charmonium hybrid states, each of which is composed of one valence charm quark and one valence charm antiquark as well as two valence gluons. We construct twenty double-gluon charmonium hybrid currents with various quantum numbers, and use them to perform QCD sum rule analyses. We refer to Ref.~\cite{Tang:2021zti} for more QCD sum rule studies. Especially, these currents can reach the exotic quantum numbers $J^{PC} =0^{--}/0^{+-}/1^{-+}/2^{+-}/3^{-+}$, whose masses are calculated to be $7.28^{+0.38}_{-0.43}$~GeV, $5.19^{+0.36}_{-0.46}$~GeV, $5.46^{+0.41}_{-0.62}$~GeV, $4.48^{+0.25}_{-0.31}$~GeV, and $5.54^{+0.35}_{-0.43}$~GeV, respectively. These mass values are accessible in the LHC experiments.

We further study their possible decay patterns from the two-/three-meson and two-baryon decay processes. Since these three processes are both at the $\mathcal{O}(\alpha_s)$ order, the three-meson and two-baryon decay patterns are generally not suppressed severely compared to the two-meson decay pattern. Especially, we propose to search for the $J^{PC}=2^{+-}/3^{-+}$ states in the $D^*\bar D^{(*)}/D^{*}_s \bar D^{(*)}_s/\Sigma_c^* \bar \Sigma_c^{(*)}/\Xi_c^* \bar \Xi_c^{(\prime,*)}$ channels directly at LHC, given that they may have relatively smaller widths due to their limited decay patterns. Experimental investigations on these states and decay channels can be useful in classifying the nature of the hybrid state, thus serving as a direct test of QCD in the low energy sector.

{\it Double-gluon charmonium hybrid currents} ---
As the first step, we combine the charm quark field $c_a(x)$, the charm antiquark field $\bar c_a(x)$, the gluon field strength tensor $G^n_{\mu\nu}(x)$, and the dual gluon field strength tensor $\tilde G^n_{\mu\nu}(x) = G^{n,\rho\sigma}(x) \times \epsilon_{\mu\nu\rho\sigma}/2$ to construct the double-gluon charmonium hybrid currents. Here $a=1\cdots3$ and $n=1\cdots8$ are color indices, and $\mu \cdots \sigma$ are Lorentz indices. These currents can be generally constructed by combining the color-octet quark-antiquark fields
\begin{gather}
\nonumber
\bar c_a \lambda_n^{ab} c_b \, , \, \bar c_a \lambda_n^{ab} \gamma_5 c_b \, ,
\\
\bar c_a \lambda_n^{ab} \gamma_\mu c_b \, , \, \bar c_a \lambda_n^{ab} \gamma_\mu \gamma_5 c_b \, , \, \bar c_a \lambda_n^{ab} \sigma_{\mu\nu} c_b \, ,
\end{gather}
and the color-octet double-gluon fields
\begin{equation}
d^{npq} G_p^{\alpha\beta} G_q^{\gamma\delta} \, , \, f^{npq} G_p^{\alpha\beta} G_q^{\gamma\delta} \, ,
\end{equation}
where $d^{npq}$ and $f^{npq}$ are the totally symmetric and antisymmetric $SU(3)$ structure constants, respectively.

In the present study we shall investigate as many as twenty double-gluon charmonium hybrid currents with various quantum numbers $J^{PC}$. We write them as $J^{\alpha_1\beta_1\cdots\alpha_J\beta_J}_{J^{PC}_{A/B}}$ or $J^{\alpha_1\cdots\alpha_J}_{J^{PC}_{C}}$, where the subscripts $A$, $B$, and $C$ denote the quark-antiquark fields $\bar c_a \lambda_n^{ab} \gamma_5 c_b$, $\bar c_a \lambda_n^{ab} \sigma_{\mu\nu} c_b$, and $\bar c_a \lambda_n^{ab} \gamma_\mu c_b$, respectively:
\begin{eqnarray}
\nonumber
J_{0^{++}_A} &=& \bar c_a \gamma_5 \lambda_n^{ab} c_b~d^{npq}~g_s^2 G_p^{\mu\nu} \tilde G_{q,\mu\nu} \, ,
\\ \nonumber
J_{0^{-+}_A} &=& \bar c_a \gamma_5 \lambda_n^{ab} c_b~d^{npq}~g_s^2 G_p^{\mu\nu} G_{q,\mu\nu} \, ,
\\ \nonumber
J^{\alpha\beta}_{1^{+-}_A} &=& \bar c_a \gamma_5 \lambda_n^{ab} c_b~f^{npq}~g_s^2 G_p^{\alpha\mu} \tilde G_{q,\mu}^\beta - \{ \alpha \leftrightarrow \beta \} \, ,
\\ \nonumber
J^{\alpha\beta}_{1^{--}_A} &=& \bar c_a \gamma_5 \lambda_n^{ab} c_b~f^{npq}~g_s^2 G_p^{\alpha\mu} G_{q,\mu}^\beta - \{ \alpha \leftrightarrow \beta \} \, ,
\\ \nonumber
J^{\alpha_1\beta_1,\alpha_2\beta_2}_{2^{++}_A} &=& \bar c_a \gamma_5 \lambda_n^{ab} c_b~d^{npq}~\mathcal{S}[ g_s^2 G_p^{\alpha_1\beta_1} \tilde G_q^{\alpha_2\beta_2} ] \, ,
\\ \nonumber
J^{\alpha_1\beta_1,\alpha_2\beta_2}_{2^{-+}_A} &=& \bar c_a \gamma_5 \lambda_n^{ab} c_b~d^{npq}~\mathcal{S}[ g_s^2 G_p^{\alpha_1\beta_1} G_q^{\alpha_2\beta_2} ] \, ,
\\ \nonumber
J^{\alpha_1\beta_1,\alpha_2\beta_2}_{2^{+-}_A} &=& \bar c_a \gamma_5 \lambda_n^{ab} c_b~f^{npq}~\mathcal{S}[ g_s^2 G_p^{\alpha_1\beta_1} \tilde G_q^{\alpha_2\beta_2} ] \, ,
\\ \nonumber
J_{0^{++}_B} &=& \bar c_a \sigma^{\mu\nu} \lambda_n^{ab} c_b~f^{npq}~g_s^2 G_{p,\nu\rho} G_{q,\mu}^\rho \, ,
\\ \nonumber
J_{0^{-+}_B} &=& \bar c_a \sigma^{\mu\nu} \lambda_n^{ab} c_b~f^{npq}~g_s^2 G_{p,\nu\rho} \tilde G_{q,\mu}^\rho \, ,
\\ \nonumber
J^{\alpha\beta}_{1^{++}_B} &=& \mathcal{S}[ \bar c_a \sigma_{\alpha_1\beta_1} \lambda_n^{ab} c_b~f^{npq} ~ g_s^2 G_{p,\alpha_2\mu} G_{q,\beta_2}^{\mu} ]
\\ && \nonumber ~~~~~ \times g^{\beta_1 \beta_2} (g^{\alpha\alpha_1}g^{\beta\alpha_2} - g^{\beta\alpha_1}g^{\alpha\alpha_2} ) \, ,
\\ \nonumber
J^{\alpha\beta}_{1^{-+}_B} &=& \mathcal{S}[ \bar c_a \sigma_{\alpha_1\beta_1} \lambda_n^{ab} c_b~f^{npq} ~ g_s^2 G_{p,\alpha_2\mu} \tilde G_{q,\beta_2}^{\mu} ]
\\ && \nonumber ~~~~~ \times  g^{\beta_1 \beta_2} (g^{\alpha\alpha_1}g^{\beta\alpha_2} - g^{\beta\alpha_1}g^{\alpha\alpha_2} ) \, ,
\\ \nonumber
J^{\alpha\beta}_{1^{+-}_B} &=& \bar c_a \sigma^{\alpha\beta} \lambda_n^{ab} c_b~d^{npq}~g_s^2 G_p^{\mu\nu} G_{q,\mu\nu} \, ,
\\ \nonumber
J^{\alpha\beta}_{1^{--}_B} &=& \bar c_a \sigma^{\alpha\beta} \lambda_n^{ab} c_b~d^{npq}~g_s^2 G_p^{\mu\nu} \tilde G_{q,\mu\nu} \, ,
\\ \nonumber
J^{\alpha_1\beta_1,\alpha_2\beta_2}_{2^{++}_B} &=& \mathcal{S}[ \bar c_a \sigma^{\alpha_1\beta_1} \lambda_n^{ab} c_b~f^{npq} ~ g_s^2 G_p^{\alpha_2\mu} G_{q,\mu}^{\beta_2} ] \, ,
\\ \nonumber
J^{\alpha_1\beta_1,\alpha_2\beta_2}_{2^{-+}_B} &=& \mathcal{S}[ \bar c_a \sigma^{\alpha_1\beta_1} \lambda_n^{ab} c_b~f^{npq} ~ g_s^2 G_p^{\alpha_2\mu} \tilde G_{q,\mu}^{\beta_2} ] \, ,
\\ \nonumber
J^{\alpha_1\beta_1\cdots\alpha_3\beta_3}_{3^{-+}_B} &=& \mathcal{S}[ \bar c_a \sigma^{\alpha_1\beta_1} \lambda_n^{ab} c_b~f^{npq}~g_s^2 G_p^{\alpha_2\beta_2} \tilde G_q^{\alpha_3\beta_3}] \, ,
\\ \nonumber
J^{\alpha_1\beta_1\cdots\alpha_3\beta_3}_{3^{+-}_B} &=& \mathcal{S}[ \bar c_a \sigma^{\alpha_1\beta_1} \lambda_n^{ab} c_b~d^{npq}~g_s^2 G_p^{\alpha_2\beta_2} G_q^{\alpha_3\beta_3}] \, ,
\\ \nonumber
J^{\alpha_1\beta_1\cdots\alpha_3\beta_3}_{3^{--}_B} &=& \mathcal{S}[ \bar c_a \sigma^{\alpha_1\beta_1} \lambda_n^{ab} c_b~d^{npq}~g_s^2 G_p^{\alpha_2\beta_2} \tilde G_q^{\alpha_3\beta_3}] \, ,
\\ \nonumber
J^\alpha_{1^{+-}_C} &=& \bar c_a \gamma^\alpha \lambda_n^{ab} c_b~d^{npq}~g_s^2 G_p^{\mu\nu} \tilde G_{q,\mu\nu} \, ,
\\
J^\alpha_{1^{--}_C} &=& \bar c_a \gamma^\alpha \lambda_n^{ab} c_b~d^{npq}~g_s^2 G_p^{\mu\nu} G_{q,\mu\nu} \, .
\label{def:currents}
\end{eqnarray}
Here $\mathcal{S}$ represents the symmetrization and subtracting trace terms in the two sets $\{\alpha_1\cdots\alpha_J\}$ and $\{\beta_1\cdots\beta_J\}$ as well as the anti-symmetrization in the sets $\{\alpha_1\beta_1\}\cdots\{\alpha_J\beta_J\}$, simultaneously.

The double-gluon hybrid currents with the light quark-antiquark fields $\bar q_a \lambda_n^{ab} \gamma_5 q_b$ and $\bar q_a \lambda_n^{ab} \sigma_{\mu\nu} q_b$ ($q=u,d,s$) have been systematically investigated in Refs.~\cite{Chen:2021smz,Su:2022fqr,Su:2023jxb}, and in the present study we just need to replace the light quark fields by the charm quark fields. However, these currents can only reach the exotic quantum numbers $J^{PC} =1^{-+}/2^{+-}/3^{-+}$, and we need the other two currents $J^\alpha_{1^{\pm-}_C}$ with the quark-antiquark field $\bar c_a \lambda_n^{ab} \gamma_\mu c_b$ in order to study the exotic quantum numbers $J^{PC} =0^{--}/0^{+-}$, as discussed below.

{\it QCD sum rule analyses} ---
The QCD sum rule method has been widely applied in the study of hadron physics~\cite{Shifman:1978bx,Reinders:1984sr,Narison:2002woh,Nielsen:2009uh}. In this letter we apply this method to study the double-gluon charmonium hybrid currents listed in Eqs.~(\ref{def:currents}). We use the current $J^\alpha_{1^{+-}_C}$ as an example and calculate its two-point correlation function
%
\begin{eqnarray}
\Pi^{\alpha\beta}(q^2) &\equiv& i \int d^4x e^{iqx} \langle 0 | {\bf T}[J^\alpha_{1^{+-}_C}(x) J^{\beta\dagger}_{1^{+-}_C}(0)] | 0 \rangle
\label{eq:correlation}
\\ \nonumber &=& (q^\alpha q^\beta - q^2 g^{\alpha\beta})~\Pi_1(q^2) + q^\alpha q^\beta \Pi_0(q^2) \, ,
\end{eqnarray}
%
at both the hadron and quark-gluon levels. The correlation functions $\Pi_1(q^2)$ and $\Pi_0(q^2)$ are respectively contributed by the $J^{PC} = 1^{+-}$ and $0^{--}$ states through
\begin{eqnarray}
\langle 0 | J^\alpha_{1^{+-}_C} | X ; 1^{+-}_C \rangle &=& \epsilon^\alpha f_{1^{+-}_C} \, ,
\\ \langle 0 | J^\alpha_{1^{+-}_C} | X ; 0^{--}_C \rangle &=& q^\alpha f_{0^{--}_C} \, .
\end{eqnarray}
We concentrate on the exotic term $\Pi_0(q^2)$ and extract its spectral density $\rho(s) \equiv {\rm Im}\Pi_0(s)/\pi$ through the dispersion relation
%
\begin{equation}
\Pi_0(q^2) = \int_{s_<}^\infty \frac{\rho(s)}{s-q^2-i\varepsilon}ds \, ,
\label{eq:rho}
\end{equation}
%
where $s_< = 4 m_c^2$ is a kinematic limit, {\it i.e.}, the square of the sum of the current charm quark masses of the hadron.

At the hadron level we parameterize $\Pi_0(q^2)$ using one-pole-dominance assumption for the possibly-existing ground state $|X ; 0^{--}_C\rangle$ and the continuum
%
\begin{eqnarray}
\nonumber q^\alpha q^\beta \rho(s) &\equiv& \sum_n\delta(s-M^2_n) \langle 0| J^\alpha_{1^{+-}_C} | n\rangle \langle n| J^{\beta\dagger}_{1^{+-}_C} |0 \rangle
\\ &=& q^\alpha q^\beta f^2_{0^{--}_C} \delta(s-M^2_X) + \rm{continuum} \, .
\end{eqnarray}
%

\begin{figure}[hbtp]
\begin{center}
\subfigure[(a)]{
\scalebox{0.1}{\includegraphics{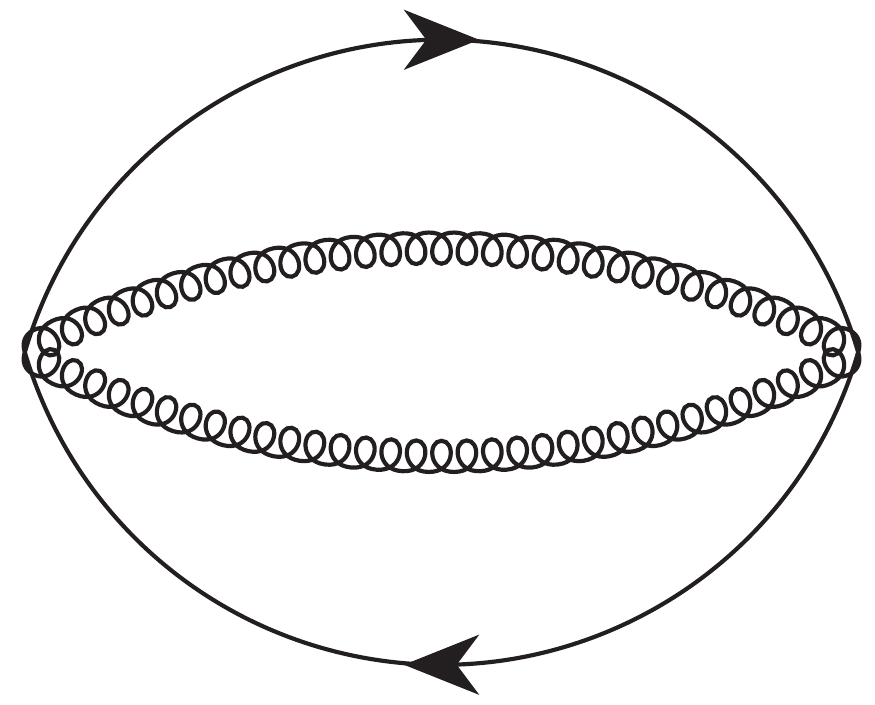}}}
\\
\subfigure[(b--1)]{
\scalebox{0.1}{\includegraphics{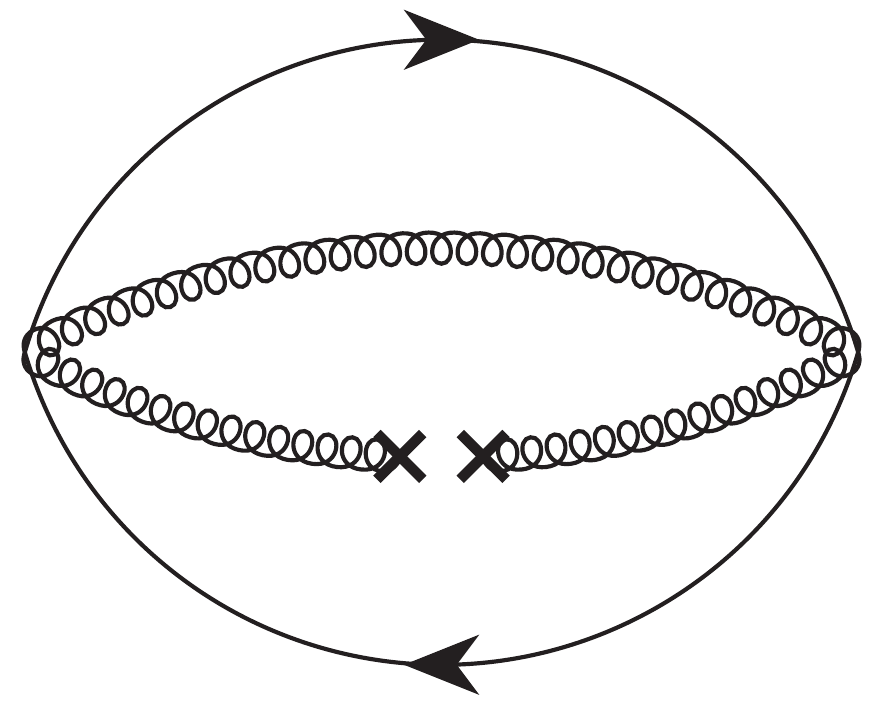}}}~~~~~~~~~~
\subfigure[(b--2)]{
\scalebox{0.1}{\includegraphics{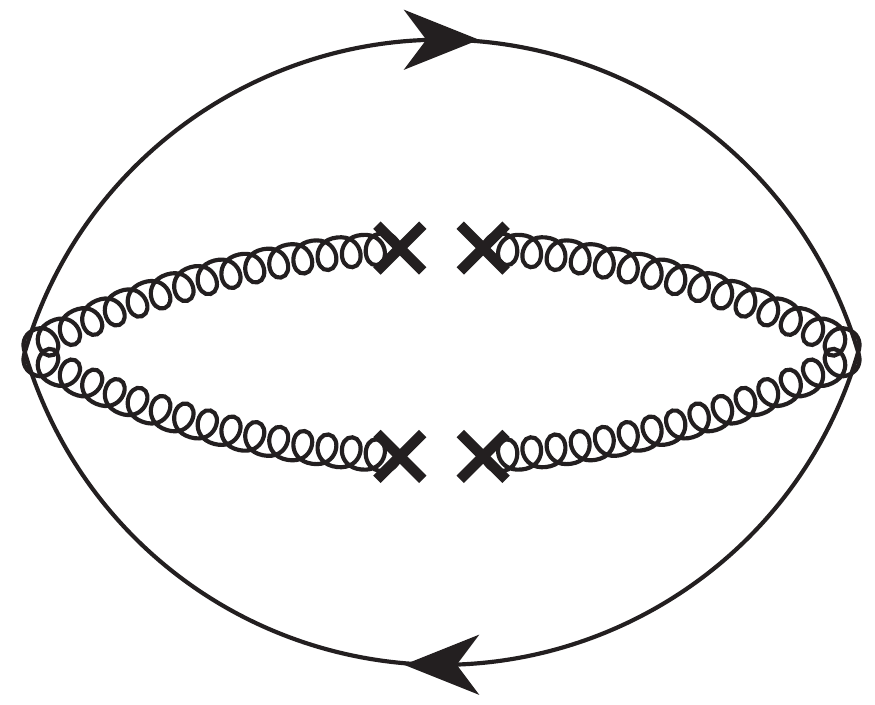}}}
\\
\subfigure[(c--1)]{
\scalebox{0.1}{\includegraphics{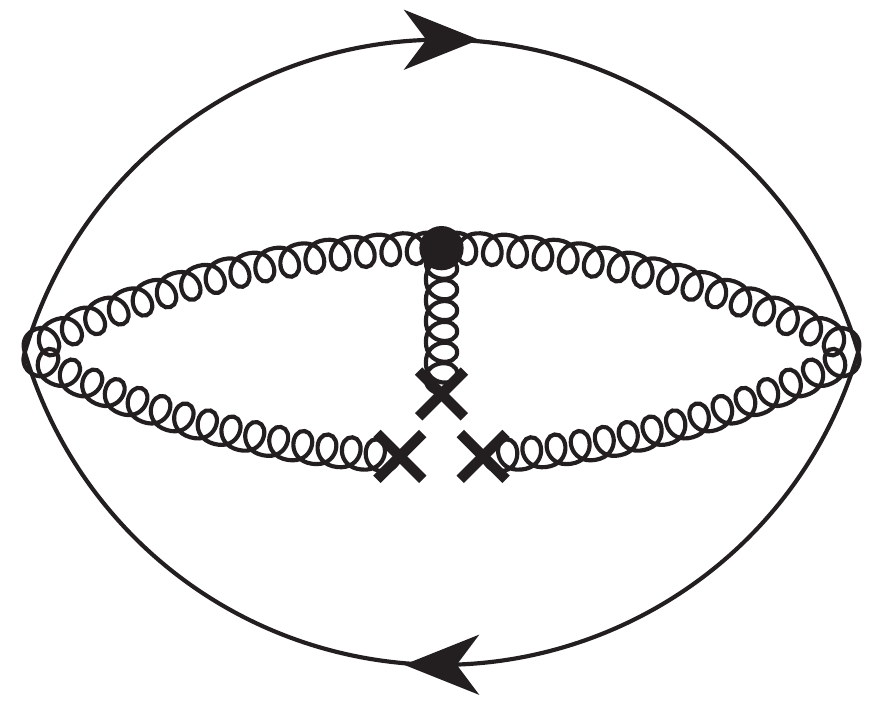}}}~
\subfigure[(c--2)]{
\scalebox{0.1}{\includegraphics{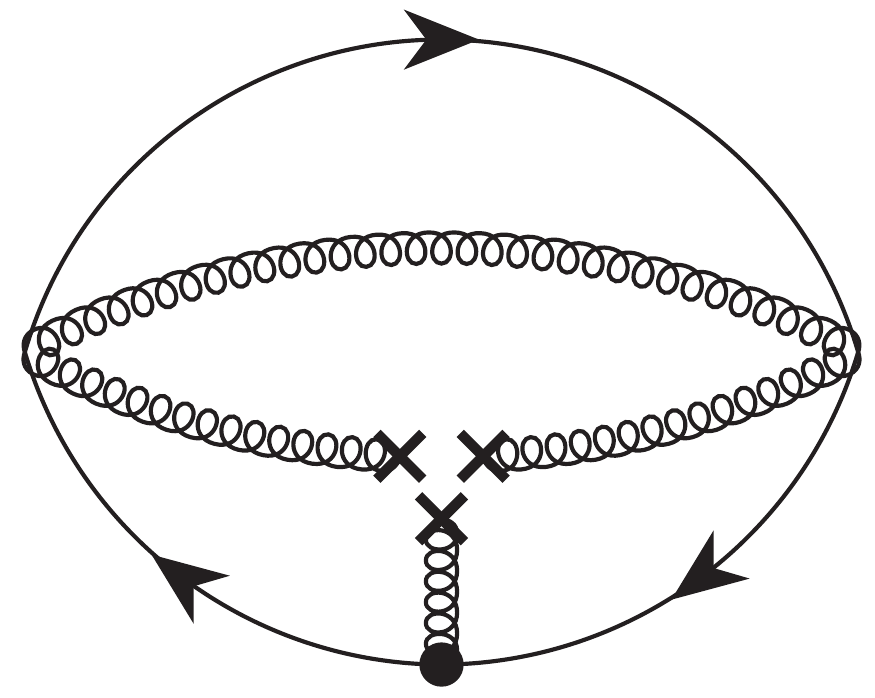}}}~
\subfigure[(c--3)]{
\scalebox{0.1}{\includegraphics{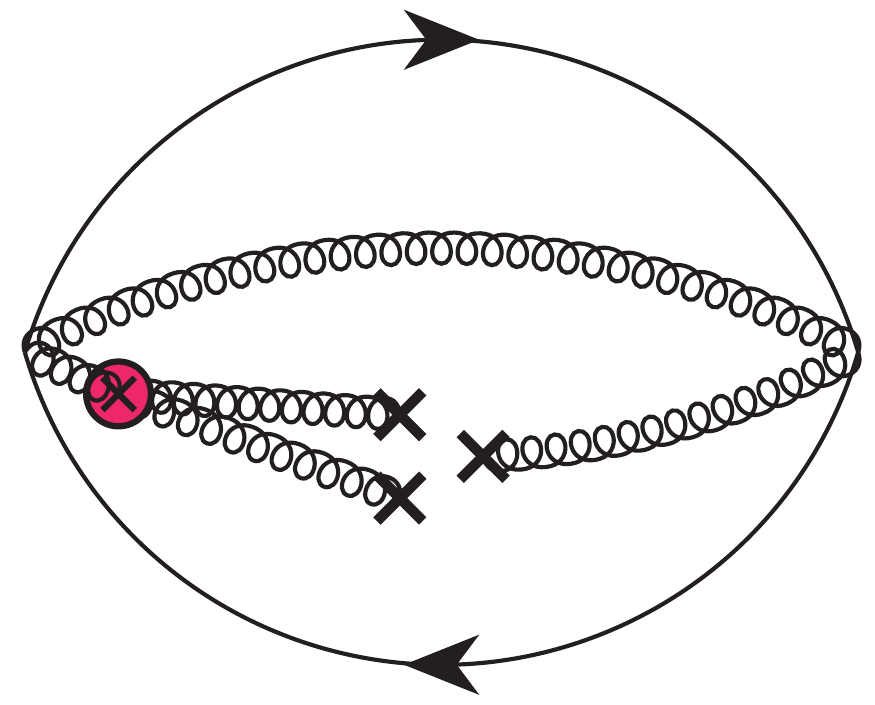}}}~
\subfigure[(c--4)]{
\scalebox{0.1}{\includegraphics{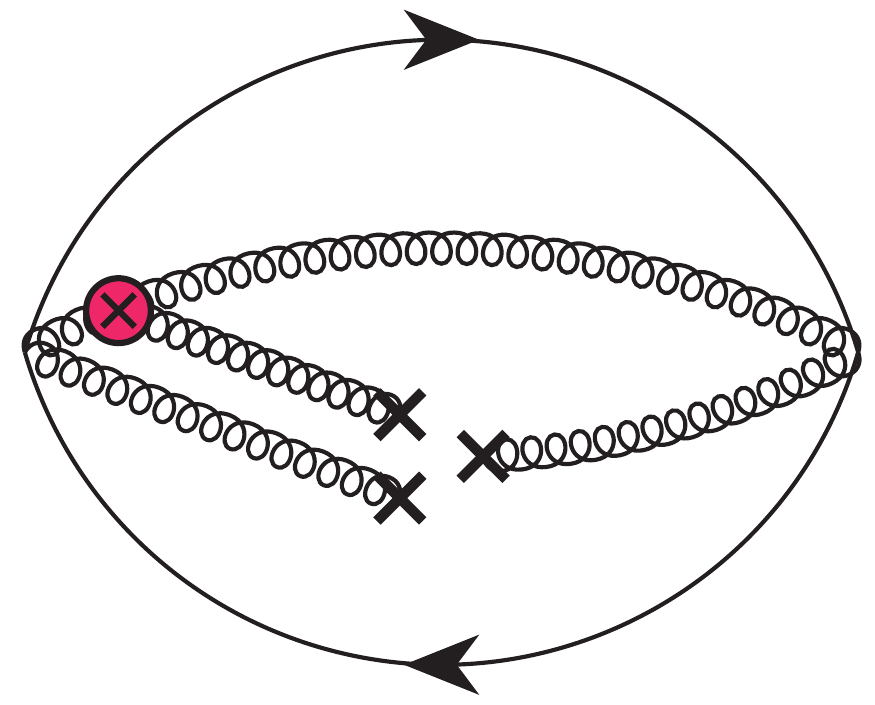}}}
\\
\subfigure[(d--1)]{
\scalebox{0.1}{\includegraphics{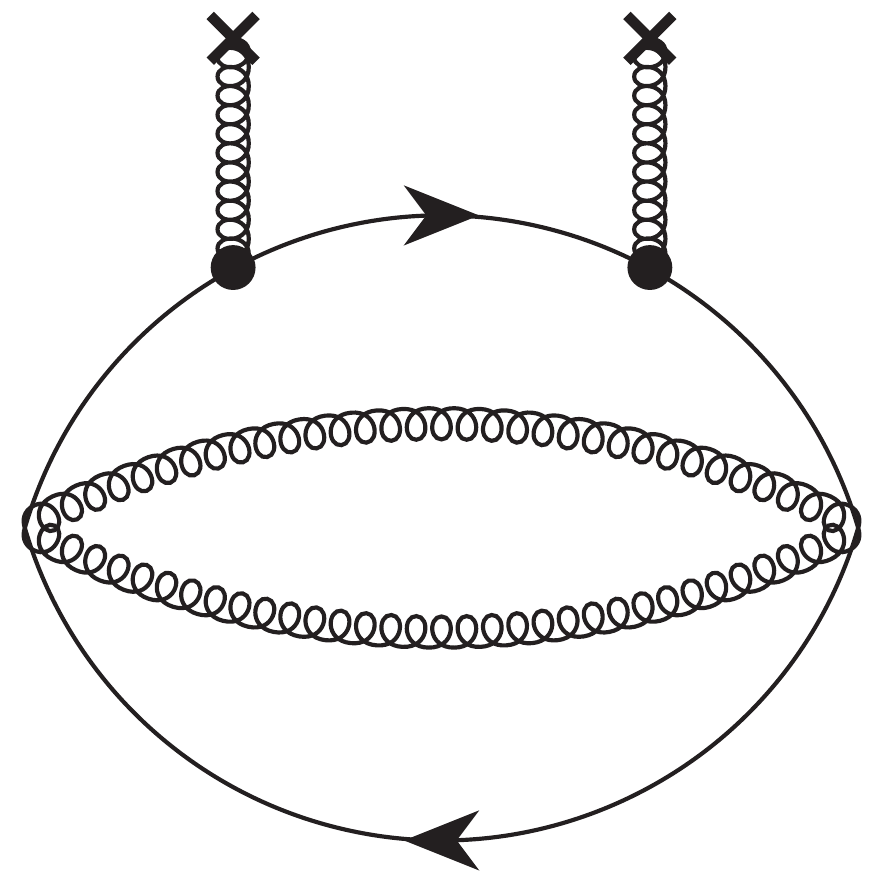}}}~
\subfigure[(d--2)]{
\scalebox{0.1}{\includegraphics{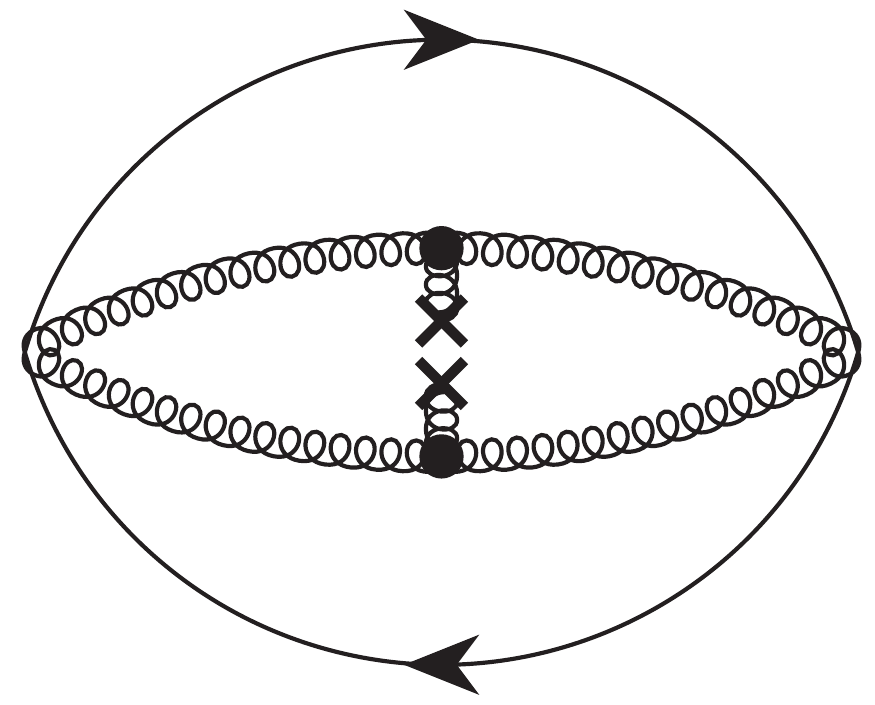}}}~
\subfigure[(d--3)]{
\scalebox{0.1}{\includegraphics{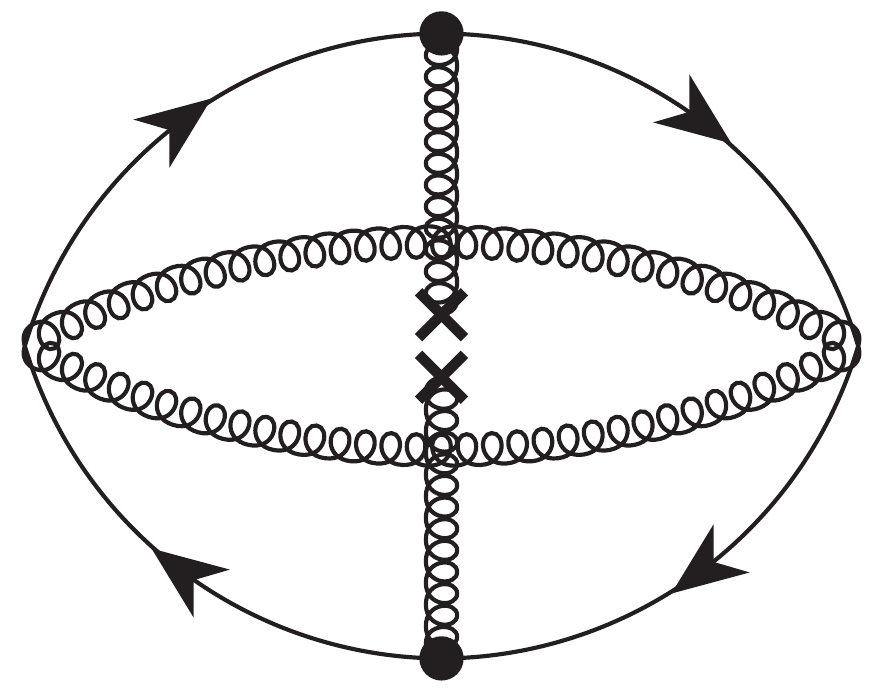}}}~
\subfigure[(d--4)]{
\scalebox{0.1}{\includegraphics{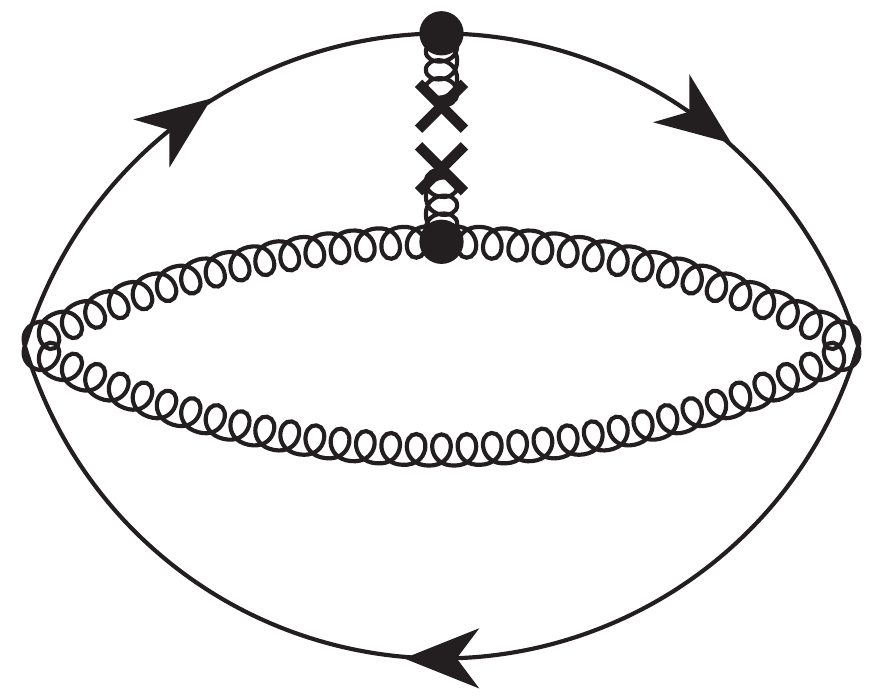}}}
\\
\subfigure[(e--1)]{
\scalebox{0.1}{\includegraphics{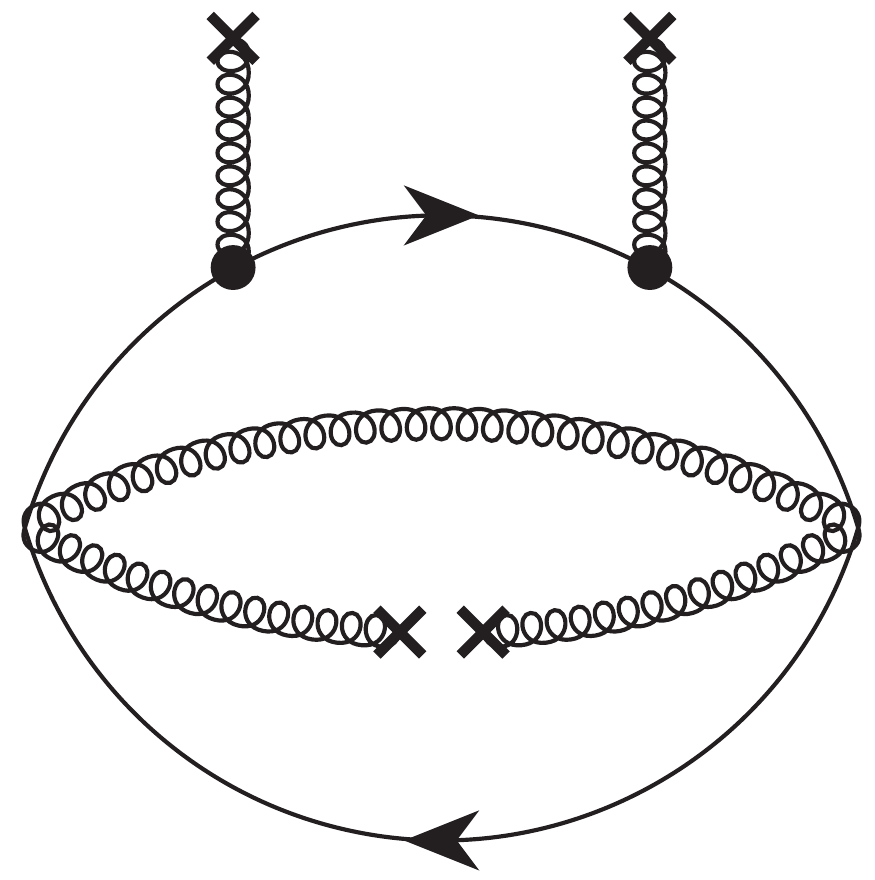}}}~~~
\subfigure[(e--2)]{
\scalebox{0.1}{\includegraphics{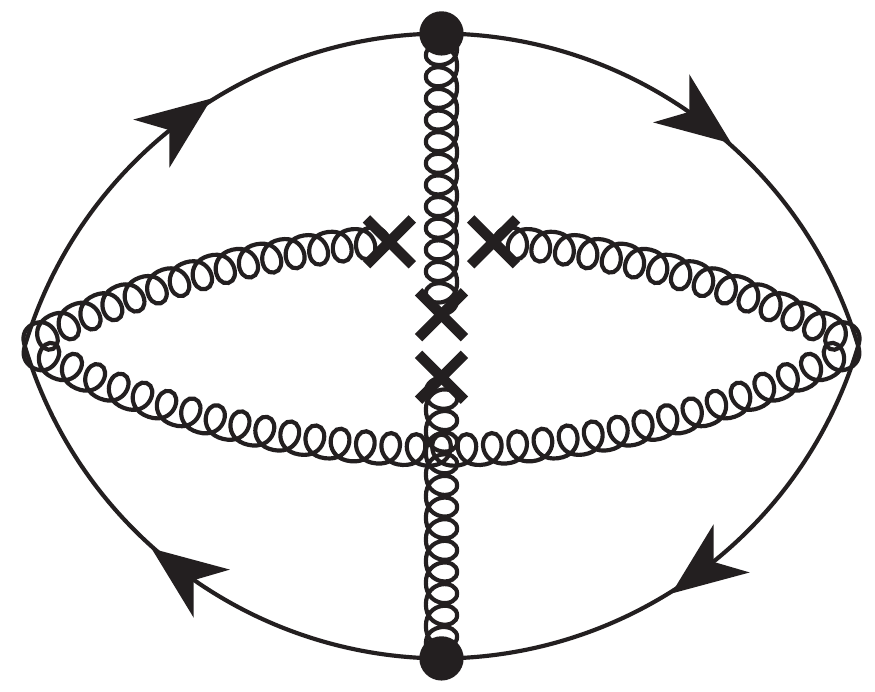}}}~~~
\subfigure[(e--3)]{
\scalebox{0.1}{\includegraphics{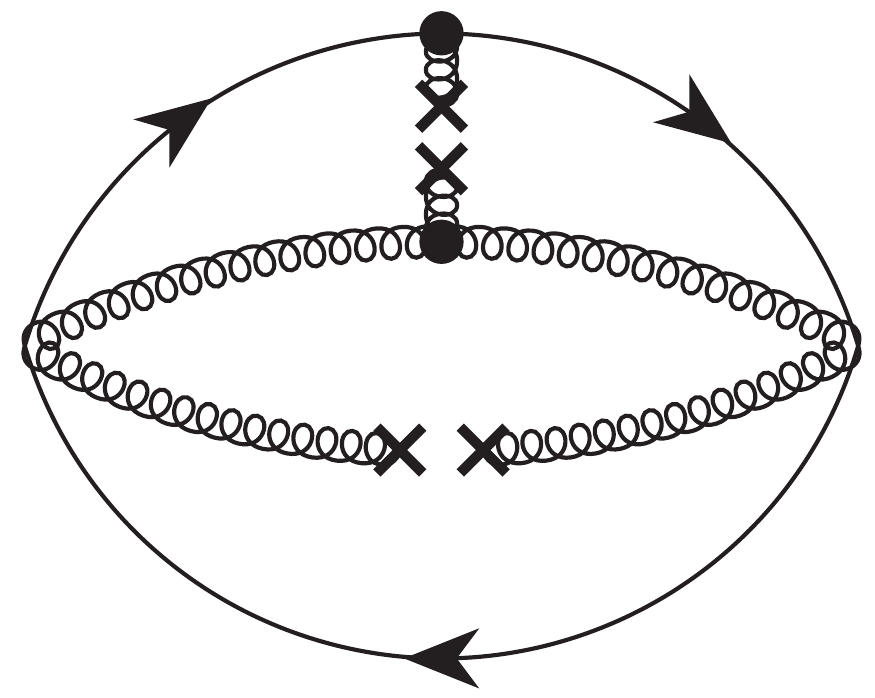}}}
\\
\subfigure[(f--1)]{
\scalebox{0.1}{\includegraphics{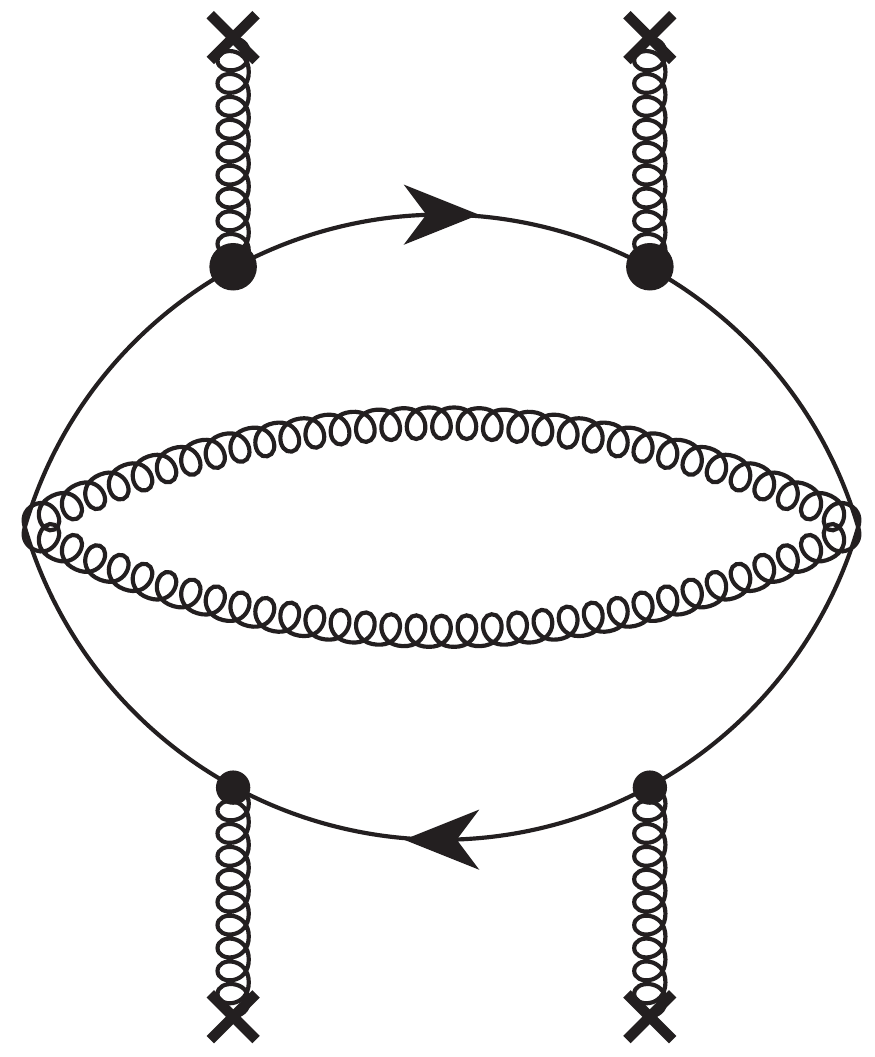}}}~~~
\subfigure[(f--2)]{
\scalebox{0.1}{\includegraphics{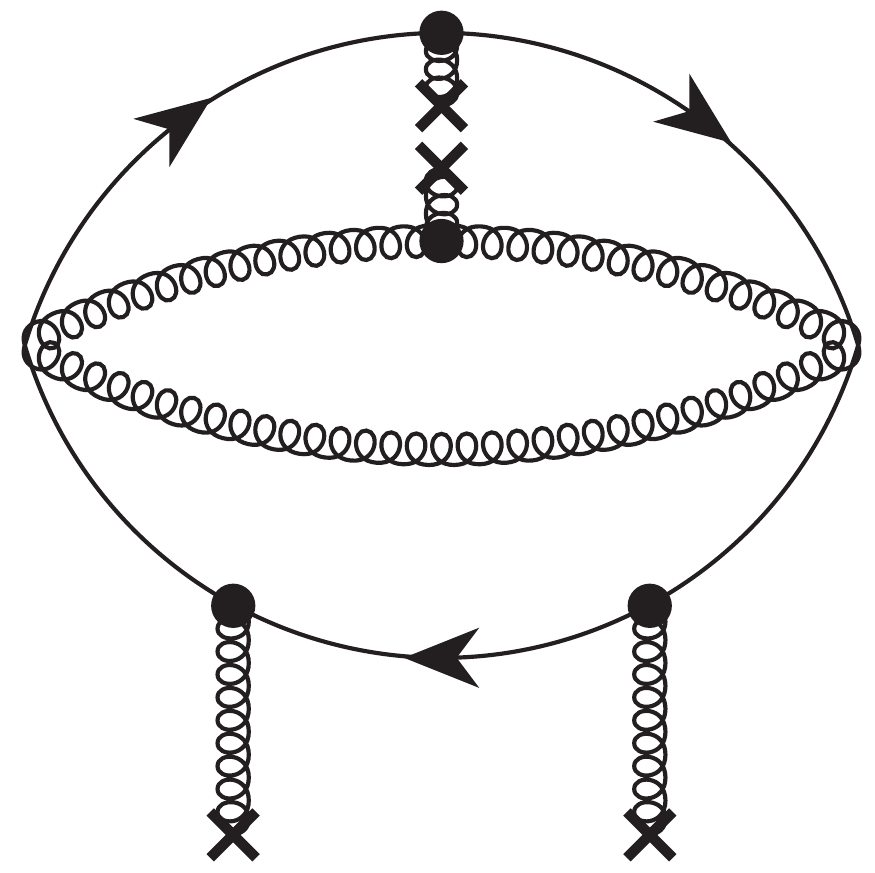}}}~~~
\subfigure[(f--3)]{
\scalebox{0.1}{\includegraphics{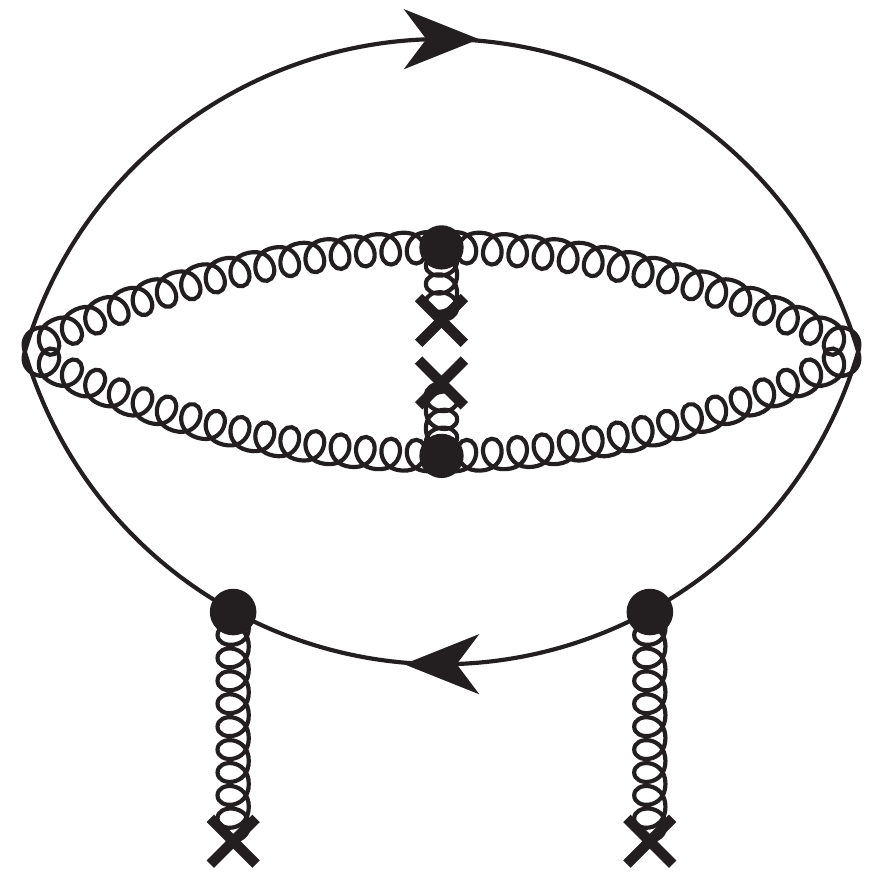}}}
\end{center}
\caption{Feynman diagrams for the double-gluon charmonium hybrid state: (a) and (b--i) are proportional to $\alpha_s^2 \times g_s^0$; (c--i) are proportional to $\alpha_s^2 \times g_s^1$; (d--i) and (e--i) are proportional to $\alpha_s^2 \times g_s^2$; (f--i) are proportional to $\alpha_s^2 \times g_s^4$.}
\label{fig:feynman}
\end{figure}

At the quark-gluon level we calculate $\Pi_0(q^2)$ and extract the spectral density $\rho(s)$ using the method of operator product expansion (OPE). The obtained results are given in the supplementary file ``OPE.nb'', with the integration parameters $\mathcal{F}(s)=m_c^2(\alpha+\beta)-\alpha\beta s$, $\mathcal{H}(s)=m_c^2-\alpha(1-\alpha) s$, $\alpha_{min}=\frac{1-\sqrt{1-4m_c^2/s}}{2}$, $\alpha_{max}=\frac{1+\sqrt{1-4m_c^2/s}}{2}$, $\beta_{min}=\frac{\alpha m_c^2}{\alpha s-m_c^2}$, and $\beta_{max}=1-\alpha$. The other spectral densities calculated in the present study are also summarized there. In the calculations we have taken into account the Feynman diagrams depicted in Fig.~\ref{fig:feynman}, and calculated $\rho(s)$ up to the dimension eight ($D=8$) condensates. We have calculated all the diagrams proportional to $\alpha_s^2 \times g_s^0$ and $\alpha_s^2 \times g_s^1$, while we have partly calculated the diagrams proportional to $\alpha_s^2 \times g_s^{n\geq2}$. The gluon field strength tensor
\begin{equation}
G^n_{\mu\nu} = \partial_\mu A_\nu^n  -  \partial_\nu A_\mu^n  +  g_s f^{npq} A_{p,\mu} A_{q,\nu} \, ,
\label{eq:pro}
\end{equation}
can be naturally separated into two parts: we use the single-gluon-line to describe the former two terms $\partial_\mu A_\nu^n  -  \partial_\nu A_\mu^n$, and we use the double-gluon-line with a red vertex to describe the third term $g_s f^{npq} A_{p,\mu} A_{q,\nu}$, {\it e.g.}, see the diagram depicted in Fig.~\ref{fig:feynman}(c--3). Eq.~(\ref{eq:pro}) indicates that there can exist a significant mixing among the single-/double-/triple-gluon hybrid states, and moreover, these hybrid states can also mix with the conventional mesons, tetraquark states, and glueballs, etc. It is still difficult to investigate this effect, so there is still a long long way to understand glueballs and hybrid states as well as the gluon degree of freedom.

After performing the Borel transformation to Eq.~(\ref{eq:rho}) at both the hadron and quark-gluon levels, we obtain
%
\begin{equation}
\Pi(s_0, M_B^2) \equiv f^2_X e^{-M_X^2/M_B^2} = \int_{s_<}^{s_0}\rho(s) e^{-s/M_B^2} ds \, ,
\label{eq:pi}
\end{equation}
%
where the continuum has been approximated as the OPE spectral density above the threshold value $s_0$. Eq.~(\ref{eq:pi}) can be used to calculate the mass of $|X ; 0^{--}_C\rangle$ through
%
\begin{equation}
M^2_X(s_0, M_B) = \frac{\int_{s_<}^{s_0} \rho(s) e^{-s/M_B^2} s ds}{\int_{s_<}^{s_0} \rho(s) e^{-s/M_B^2} ds} \, .
\label{eq:mass}
\end{equation}
%

{\it Numerical analyses} ---
We perform numerical analyses using the following values for various QCD parameters at the QCD scale $\Lambda_{\rm QCD} = 300$~MeV and the renormalization scale $2$~GeV~\cite{pdg,Narison:2011xe,Narison:2018dcr}:
%
\begin{eqnarray}
\nonumber \alpha_s(Q^2) &=& {4\pi \over 11 \ln(Q^2/\Lambda_{\rm QCD}^2)} \, ,
\\ m_{c}(m_{c}) &=&1.27 \pm 0.02 {\rm~GeV} \, ,
\label{eq:condensate}
\\ \nonumber \langle \alpha_s GG\rangle &=& (6.35 \pm 0.35) \times 10^{-2} \mbox{ GeV}^4 \, ,
\\ \nonumber \langle g_s^3G^3\rangle &=& (8.2 \pm 1.0) \times \langle \alpha_s GG\rangle  \mbox{ GeV}^2 \, .
\end{eqnarray}
%

\begin{figure}[]
\begin{center}
\subfigure[(a)]{\includegraphics[width=0.22\textwidth]{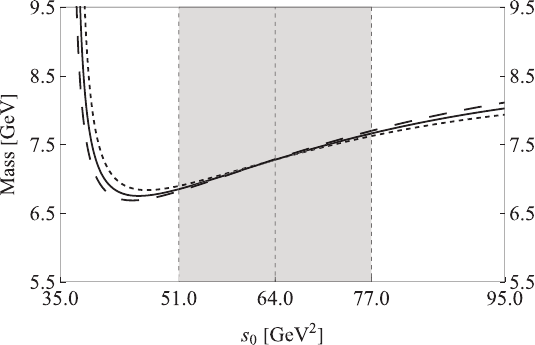}}
~~~
\subfigure[(b)]{\includegraphics[width=0.22\textwidth]{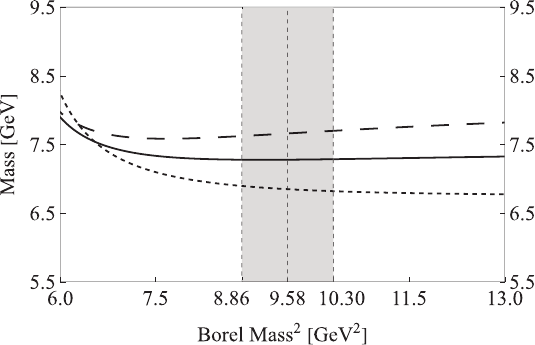}}
\caption{Mass of the double-gluon hybrid state $|X ; 0^{--}_C\rangle$ with respect to (a) the threshold value $s_0$ and (b) the Borel mass $M_B$. In the subfigure (a) the short-dashed/middle-dashed/long-dashed curves are obtained by setting $M_B^2 = 8.86/9.58/10.30$~GeV$^2$, respectively. In the subfigure (b) the short-dashed/middle-dashed/long-dashed curves are obtained by setting $s_0 = 51.0/64.0/77.0$~GeV$^2$, respectively.}
\label{fig:mass}
\end{center}
\end{figure}

As shown in Eq.~(\ref{eq:mass}), the mass of $|X ; 0^{--}_C\rangle$ depends on two free parameters: the Borel mass $M_B$ and the threshold value $s_0$. Firstly, we investigate the OPE convergence by requiring a) the $\alpha_s^2 \times g_s^{n\geq2}$ terms to be less than 5\%, b) the $D=8$ terms to be less than 10\%, and c) the $D=6$ terms to be less than 20\%:
\begin{eqnarray}
\mbox{CVG}_A &\equiv& \left|\frac{ \Pi^{g_s^{n\geq6}}(s_0, M_B^2) }{ \Pi(s_0, M_B^2) }\right| \leq 5\% \, ,
\\
\mbox{CVG}_B &\equiv& \left|\frac{ \Pi^{{\rm D=8}}(s_0, M_B^2) }{ \Pi(s_0, M_B^2) }\right| \leq 10\% \, ,
\\
\mbox{CVG}_C &\equiv& \left|\frac{ \Pi^{{\rm D=6}}(s_0, M_B^2) }{ \Pi(s_0, M_B^2) }\right| \leq 20\% \, .
\end{eqnarray}
Secondly, we investigate the one-pole-dominance assumption by requiring the pole contribution (PC) to be larger than 40\%:
\begin{equation}
\mbox{PC} \equiv \left|\frac{ \Pi(s_0, M_B^2) }{ \Pi(\infty, M_B^2) }\right| \geq 40\% \, .
\end{equation}
Altogether, we determine the Borel window to be $8.86$~GeV$^2 \leq M_B^2 \leq 10.30$~GeV$^2$ when setting $s_0 = 64.0$~GeV$^2$. We redo the same procedures and find that there exist the Borel windows as long as $s_0 \geq s_0^{\rm min} = 58.3$~GeV$^2$. Accordingly, we set $s_0$ to be slightly larger and determine the working regions to be $51.0$~GeV$^2 \leq s_0 \leq 77.0$~GeV$^2$ and $8.86$~GeV$^2 \leq M_B^2 \leq 10.30$~GeV$^2$, where the mass of $|X ; 0^{--}_C\rangle$ is calculated to be
\begin{equation}
M_{|X ; 0^{--}_C\rangle} = 7.28^{+0.38}_{-0.43}{\rm~GeV} \, .
\end{equation}
Its uncertainty comes from the threshold value $s_0$, the Borel mass $M_B$, and various QCD parameters listed in Eqs.~(\ref{eq:condensate}). We show $M_{|X ; 0^{--}_C\rangle}$ in Fig.~\ref{fig:mass} with respect to $s_0$ and $M_B$. As shown in Fig.~\ref{fig:mass}(a), we find a mass minimum around $s_0 \sim 45$~GeV$^2$, and the mass dependence on $s_0$ is moderate and acceptable inside the region $51.0$~GeV$^2 \leq s_0 \leq 77.0$~GeV$^2$. As shown in Fig.~\ref{fig:mass}(b), the mass dependence on $M_B$ is weak inside the Borel window $8.86$~GeV$^2 < M_B^2 < 10.30$~GeV$^2$.

Similarly, we apply the QCD sum rule method to study the other nineteen double-gluon charmonium hybrid currents listed in Eqs.~(\ref{def:currents}). The obtained results are summarized in Table~\ref{tab:results}.

\begin{table}[hbtp]
\begin{center}
\renewcommand{\arraystretch}{1.4}
\caption{QCD sum rule results for the double-gluon charmonium hybrid states with various quantum numbers $J^{PC}$. The results with the subscripts $A$, $B$, and $C$ are extracted from the double-gluon charmonium hybrid currents with the quark-antiquark fields $\bar c_a \lambda_n^{ab} \gamma_5 c_b$, $\bar c_a \lambda_n^{ab} \sigma_{\mu\nu} c_b$, and $\bar c_a \lambda_n^{ab} \gamma_\mu c_b$, respectively.}
\begin{tabular}{c c c c c c}
\hline\hline
 & & \multicolumn{2}{c}{Working Regions} & &
\\ \cline{3-4}
$J^{PC}$ & $s_0^{min}[{\rm GeV}^2]$ & $M_B^2[{\rm GeV}^2]$ & $s_0[{\rm GeV}^2]$ & Pole[\%] & Mass[GeV]
\\ \hline
$0^{++}_A$     &   $56.3$   &  $7.89$--$9.36$   &  $62\pm12$  &  $40$--$54$  &  $7.29^{+0.33}_{-0.26}$
\\
$0^{++}_B$     &   $48.0$   &  $6.38$--$7.63$   &  $53\pm11$  &  $40$--$54$  &  $6.76^{+0.32}_{-0.24}$
\\
$0^{-+}_A$     &   $41.2$   &  $6.73$--$7.30$   &  $45\pm9$   &  $40$--$49$  &  $5.70^{+0.43}_{-0.57}$
\\
$0^{-+}_B$     &   $39.7$   &  $5.47$--$6.26$   &  $44\pm9$   &  $40$--$53$  &  $5.87^{+0.38}_{-0.56}$
\\
$0^{+-}_C$     &   $32.9$   &  $5.03$--$5.53$   &  $36\pm7$   &  $40$--$50$  &  $5.19^{+0.36}_{-0.46}$
\\
$0^{--}_C$     &   $58.3$   &  $8.86$--$10.30$  &  $64\pm13$  &  $40$--$52$  &  $7.28^{+0.38}_{-0.43}$
\\
$1^{++}_B$     &   $47.5$   &  $6.29$--$7.47$   &  $52\pm10$  &  $40$--$53$  &  $6.74^{+0.30}_{-0.18}$
\\
$1^{-+}_B$     &   $36.3$   &  $5.18$--$5.77$   &  $40\pm8$   &  $40$--$51$  &  $5.46^{+0.41}_{-0.62}$
\\
$1^{+-}_A$     &   $49.4$   &  $6.70$--$8.01$   &  $54\pm11$  &  $40$--$53$  &  $6.92^{+0.32}_{-0.12}$
\\
$1^{+-}_B$     &   $34.2$   &  $5.28$--$5.72$   &  $38\pm8$   &  $40$--$49$  &  $5.15^{+0.44}_{-0.54}$
\\
$1^{+-}_C$     &   $55.1$   &  $7.73$--$9.26$   &  $61\pm12$  &  $40$--$54$  &  $7.22^{+0.33}_{-0.27}$
\\
$1^{--}_A$     &   $37.7$   &  $6.13$--$6.61$   &  $41\pm8$   &  $40$--$48$  &  $5.39^{+0.43}_{-0.56}$
\\
$1^{--}_B$     &   $59.1$   &  $7.67$--$9.11$   &  $65\pm13$  &  $40$--$54$  &  $7.48^{+0.37}_{-0.38}$
\\
$1^{--}_C$     &   $42.0$   &  $6.83$--$7.46$   &  $46\pm9$   &  $40$--$49$  &  $5.78^{+0.43}_{-0.56}$
\\
$2^{++}_A$     &   $34.1$   &  $7.08$--$7.68$   & $38\pm8$    &  $40$--$48$  &  $5.12^{+0.39}_{-0.48}$
\\
$2^{++}_B$     &   $39.2$   &  $5.76$--$6.28$   & $43\pm9$    &  $40$--$50$  &  $5.57^{+0.45}_{-0.62}$
\\
$2^{+-}_A$     &   $23.1$   &  $3.52$--$3.89$   & $25\pm5$    &  $40$--$49$  &  $4.48^{+0.25}_{-0.31}$
\\
$2^{-+}_A$     &    $31.2$  &  $5.74$--$6.23$   & $34\pm7$    &  $40$--$48$  &  $4.97^{+0.36}_{-0.46}$
\\
$2^{-+}_B$     &    $55.8$  &  $7.16$--$8.46$   & $61\pm12$   &  $40$--$53$  &  $7.29^{+0.33}_{-0.22}$
\\
$3^{-+}_B$     &    $35.9$  &  $5.18$--$5.71$   & $39\pm8$    &  $40$--$50$  &  $5.54^{+0.35}_{-0.43}$
\\
$3^{+-}_B$     &    $36.9$  &  $6.02$--$6.71$   & $41\pm8$    &  $40$--$51$  &  $5.56^{+0.36}_{-0.45}$
\\
$3^{--}_B$     &    $32.7$  &  $6.14$--$6.64$   & $36\pm7$    &  $40$--$48$  &  $5.07^{+0.35}_{-0.43}$
\\ \hline\hline
\end{tabular}
\label{tab:results}
\end{center}
\end{table}

\begin{figure}[]
\begin{center}
\subfigure[(a)]{\scalebox{0.2}{\includegraphics{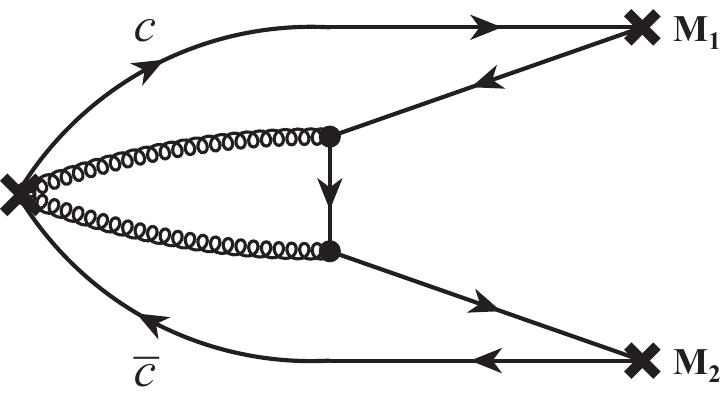}}}
~~~
\subfigure[(b)]{\scalebox{0.2}{\includegraphics{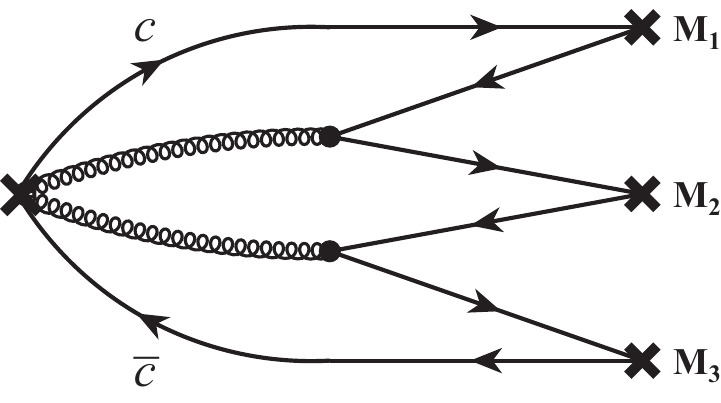}}}
~~~
\subfigure[(c)]{\scalebox{0.2}{\includegraphics{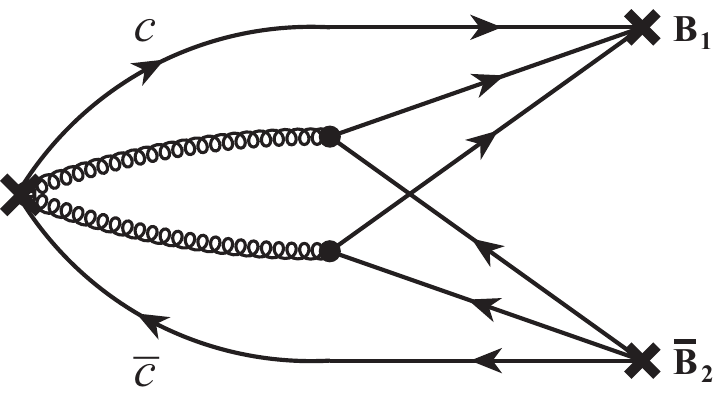}}}
\end{center}
\caption{Possible two-/three-meson and two-baryon decay processes of the double-gluon charmonium hybrid state.}
\label{fig:decay}
\end{figure}

{\it Decay analyses} ---
As depicted in Fig.~\ref{fig:decay}, the double-gluon charmonium hybrid states can decay after exciting two $\bar q q$ ($q=u,d,s$) pairs from two gluons, followed by recombining three color-octet $\bar c c/\bar q q$ pairs into two/three color-singlet mesons or two color-singlet baryons:
\begin{eqnarray}
(\bar c c)_{\mathbf{8}_C} \times (\bar q q)_{\mathbf{8}_C} &\rightarrow& (\bar c q)_{\mathbf{1}_C} (\bar q c)_{\mathbf{1}_C} \, ,
\\
(\bar c c)_{\mathbf{8}_C} \times (\bar q q)^2_{\mathbf{8}_C} &\rightarrow& (\bar q q)_{\mathbf{1}_C} (\bar c q)_{\mathbf{1}_C} (\bar q c)_{\mathbf{1}_C} \, ,
\\
(\bar c c)_{\mathbf{8}_C} \times (\bar q q)^2_{\mathbf{8}_C} &\rightarrow& (\bar c \bar q \bar q)_{\mathbf{1}_C} (c q q)_{\mathbf{1}_C} \, .
\end{eqnarray}
These three decay processes are both at the $\mathcal{O}(\alpha_s)$ order, so the three-meson and two-baryon decay patterns are generally not suppressed severely compared to the two-meson decay patterns. Comparatively speaking, their decays into one charmonium meson and light mesons are at the $\mathcal{O}(\alpha_s^2)$ order and so suppressed, but these channels can be observed in experiments more easily, such as $J/\psi \pi \pi$, $J/\psi \pi \pi \pi$, and $J/\psi K \bar K$, etc. We list in Table~\ref{tab:decay} possible $S$-wave and $P$-wave as well as several $D$-wave decay patterns of the double-gluon charmonium hybrid states with the exotic quantum numbers $J^{PC} =0^{--}/0^{+-}/1^{-+}/2^{+-}/3^{-+}$, separately for the two-/three-meson and two-baryon decay processes.

\begin{table}[hbtp]
\begin{center}
\renewcommand{\arraystretch}{1.4}
\caption{Possible $S$-wave ({\color{red}red}) and $P$-wave ({\color{blue}blue}) as well as several $D$-wave ({\color{green}green}) decay patterns of the double-gluon charmonium hybrid states with the exotic quantum numbers $J^{PC} =0^{--}/0^{+-}/1^{-+}/2^{+-}/3^{-+}$, separately for the two-/three-meson and two-baryon decay processes. Some charge-conjugated decay patterns are omitted for simplicity.}
\begin{tabular}{ c  c  c }
\hline\hline
$J^{PC}$  & ~~~Two-Meson~~~                                                                   & Three-Meson
\\ \hline
$0^{--}$  & {\color{blue} $D^*\bar D, D^{*}_s \bar D_s$}                                      & {\color{red} $D\bar D \pi/\eta^{(\prime)}, D^*\bar D^* \pi/\eta^{(\prime)}, D^*\bar D^{(*)} \rho/\omega$}
\\        & {\color{green} $\Sigma_c^* \bar \Sigma_c, \Xi_c^* \bar \Xi_c^{(\prime)}$}         & {\color{red} $D_s \bar D_s \eta^{(\prime)}, D_s^* \bar D_s^* \eta^{(\prime)}, D_s^* \bar D_s^{(*)} \phi$}
\\                                                                                           && {\color{red} $D \bar D_s K, D^* \bar D_s^* K, D \bar D_s^* K^*, D^* \bar D_s^{(*)} K^*$}
\\
$0^{+-}$  & {\color{blue}  $\Sigma_c^* \bar \Sigma_c, \Xi_c^* \bar \Xi_c^{(\prime)}$}         & {\color{blue} $D^*\bar D^{(*)} \pi/\eta^{(\prime)}, D^{(*)}\bar D^{(*)} \rho/\omega$}
\\        &                                                                                   & {\color{blue} $D_s^* \bar D_s^{(*)} \eta^{(\prime)}, D_s^{(*)} \bar D_s^{(*)} \phi$}
\\                                                                                           && {\color{blue} $D \bar D_s^* K, D^* \bar D_s^{(*)} K, D^{(*)} \bar D_s^{(*)} K^*$}
\\
$1^{-+}$  & {\color{blue} $D^* \bar D^{(*)}, D_s^* \bar D_s^{(*)}$}                           & {\color{red} $D^*\bar D^{(*)} \pi/\eta^{(\prime)},D^{(*)}\bar D^{(*)} \rho/\omega$}
\\        & {\color{red} $\Sigma_c^* \bar \Sigma_c, \Xi_c^* \bar \Xi_c^{(\prime)}$}           & {\color{red} $D_s^* \bar D_s^{(*)} \eta^{(\prime)}, D_s^{(*)} \bar D_s^{(*)} \phi$}
\\        & {\color{green} $\Sigma_c^* \bar \Sigma_c^*, \Xi_c^* \bar \Xi_c^*$}                & {\color{red} $D \bar D_s^* K, D^* \bar D_s^{(*)} K, D^{(*)} \bar D_s^{(*)} K^*$}

\\
$2^{+-}$  & {\color{green} $D^*\bar D^{(*)}, D^{*}_s \bar D^{(*)}_s$}                         & {\color{blue} $D^*\bar D^{(*)} \pi/\eta, D \bar D^{(*)} \rho/\omega, D_s \bar D_s^{*} \eta$}
\\
                                                                                             && {\color{blue} $D \bar D_s^* K, D^* \bar D_s^{(*)} K, D \bar D_s K^*$}
\\
$3^{-+}$  & {\color{green} $\Sigma_c^* \bar \Sigma_c^{(*)}, \Xi_c^* \bar \Xi_c^{(\prime,*)}$} & {\color{red} $D^*\bar D^* \rho/\omega, D^{*}_s \bar D^{*}_s\phi, D^{*} \bar D_s^{*} K^*$}
\\ \hline\hline
\end{tabular}
\label{tab:decay}
\end{center}
\end{table}

{\it Summary } ---
In this letter we study the double-gluon charmonium hybrid states with various quantum numbers. We construct twenty double-gluon charmonium hybrid currents and use them to perform QCD sum rule analyses. These currents can reach the exotic quantum numbers $J^{PC} =0^{--}/0^{+-}/1^{-+}/2^{+-}/3^{-+}$ that the conventional $\bar q q$ mesons can not reach, from which we obtain
\begin{eqnarray}
\nonumber M_{|X;0^{--}\rangle} &=& 7.28^{+0.38}_{-0.43}{\rm~GeV} \, ,
\non
M_{|X;0^{+-}\rangle} &=& 5.19^{+0.36}_{-0.46}{\rm~GeV}\, ,
\non
M_{|X;1^{-+}\rangle} &=& 5.46^{+0.41}_{-0.62}{\rm~GeV}\, ,
\non
M_{|X;2^{+-}\rangle} &=& 4.48^{+0.25}_{-0.31}{\rm~GeV}\, ,
\\
M_{|X;3^{-+}\rangle} &=& 5.54^{+0.35}_{-0.43}{\rm~GeV}\, .
\end{eqnarray}
The above mass values are accessible in the LHC experiments.

We further study possible decay patterns of the double-gluon charmonium hybrid states with the exotic quantum numbers $J^{PC} =0^{--}/0^{+-}/1^{-+}/2^{+-}/3^{-+}$, separately for the two-/three-meson and two-baryon decay processes. We propose to search for them experimentally in their possible decay channels $D^{(*)}\bar D^{(*)}(\pi/\eta/\eta^\prime/\rho/\omega)$, $D^{(*)}_s \bar D^{(*)}_s(\eta/\eta^\prime/\phi)$, $D^{(*)} \bar D^{(*)}_sK^{(*)}$, and $\Lambda_c \bar \Lambda_c/\Sigma_c^{(*)} \bar \Sigma_c^{(*)}/\Xi_c^{(\prime,*)} \bar \Xi_c^{(\prime,*)}/\Omega_c^{(*)} \bar \Omega_c^{(*)}$ etc. Especially, the $J^{PC}=2^{+-}/3^{-+}$ states may have relatively smaller widths due to their limited decay patterns, so we propose to search for them in the $D^*\bar D^{(*)}/D^{*}_s \bar D^{(*)}_s/\Sigma_c^* \bar \Sigma_c^{(*)}/\Xi_c^* \bar \Xi_c^{(\prime,*)}$ channels directly at LHC. Experimental investigations on these states and decay channels can be useful in classifying the nature of the hybrid state, thus serving as a direct test of QCD in the low energy sector.

\vspace{0.5cm}

\begin{acknowledgments}
This project is supported by
the National Natural Science Foundation of China under Grant No.~11975033, No.~12075019, No.~12175318, and No.~12070131001,
the National Key R$\&$D Program of China under Contracts No.~2020YFA0406400,
the Jiangsu Provincial Double-Innovation Program under Grant No.~JSSCRC2021488,
and
the Fundamental Research Funds for the Central Universities.
\end{acknowledgments}

\bibliographystyle{elsarticle-num}
\bibliography{ref}

\end{document}